\documentclass[twocolumn,preprintnumbers,amsmath,amssymbm,prl]{revtex4}
\usepackage{epsfig}
\usepackage{graphicx}

\begin{document}
\title{Weak Cosmic Censorship: As Strong As Ever}
\author{Shahar Hod}
\address{The Ruppin Academic Center, Emeq Hefer 40250, Israel}
\address{ }
\address{The Hadassah Institute, Jerusalem 91010, Israel}
\date{\today}

\begin{abstract}
\ \ \ Spacetime singularities that arise in gravitational collapse
are always hidden inside of black holes. This is the essence of the
weak cosmic censorship conjecture. The hypothesis, put forward by
Penrose forty years ago, is still one of the most important open
questions in general relativity. In this essay we reanalyze extreme
situations which have been considered as counterexamples to the weak
cosmic censorship conjecture. In particular, we consider the
absorption of scalar particles with large angular momentum by a
black hole. Ignoring backreaction effects may lead one to conclude
that the incident wave may over spin the black hole, thereby
exposing its inner singularity to distant observers. However, we
show that when backreaction effects are properly taken into account,
the stability of the black-hole event horizon is irrefutable. We
therefore conclude that cosmic censorship is actually respected in
this type of gedanken experiments.
\end{abstract}
\bigskip
\maketitle


The influential theorems of Hawking and Penrose \cite{HawPen}
demonstrate that spacetime singularities are ubiquitous features of
general relativity, Einstein's theory of gravity. This implies that
general relativity itself predicts its own failure to describe the
physics of these extreme situations. Nevertheless, the utility of
general relativity in describing gravitational phenomena is
maintained by the cosmic censorship conjecture
\cite{Pen,Haw1,Brady}. The weak cosmic censorship conjecture (WCCC)
asserts that spacetime singularities that arise in gravitational
collapse are always hidden inside of black holes. This statement is
based on the common wisdom that singularities are not pervasive
\cite{Brady}.

The validity of the WCCC is a necessary condition to ensure the
predictability of the laws of physics \cite{Pen,Haw1,Brady}. The
conjecture, which is widely believed to be true, has become one of
the cornerstones of general relativity. Moreover, it is being
envisaged as a basic principle of nature. However, despite the
flurry of research over the years, the validity of this conjecture
is still an open question (see e.g.
\cite{Wald1,Sin,Clar,Vis,Price,Wald2,His,KayWal,BekRos,Hub,QuiWal,Hod1,HodPir,Hod2,ForRom1,ForRom2,MatSil}
and references therein).

The destruction of a black-hole event horizon is ruled out by this
principle because it would expose the inner singularities to distant
observers. Moreover, the horizon area of a black hole, $A$, is
associated with an entropy $S_{BH}=A/4$ \cite{Beken1} (we use
natural units in which $G=c=\hbar=1$). Therefore, without any
obvious physical mechanism to compensate for the loss of the
black-hole enormous entropy, the destruction of the black-hole event
horizon would violate the generalized second law (GSL) of
thermodynamics \cite{Beken1}. For these two reasons, any process
which seems, at first sight, to remove the black-hole horizon is
expected to be unphysical. For the advocates of the cosmic
censorship principle the task remains to find out how such candidate
processes eventually fail to remove the horizon.

According to the uniqueness theorems \cite{un1,un2,un3,un4,un5}, all
stationary solutions of the Einstein-Maxwell equations are uniquely
described by the Kerr-Newman metric which is characterized by three
conserved parameters: the gravitational mass $M$, the angular
momentum $J$, and the electric charge $Q$. A black-hole solution
must satisfy the relation

\begin{equation}\label{Eq1}
M^2-Q^2-a^2 \geq 0\  ,
\end{equation}
where $a\equiv J/M$ is the specific angular momentum of the black
hole. Extreme black holes are the ones which saturate the relation
(\ref{Eq1}). As is well known, the Kerr-Newman metric with
$M^2-Q^2-a^2<0$ does not contain an event horizon, and it is
therefore associated with a naked singularity rather than a black
hole. In this work we inquire into the physical mechanism which
protects the black-hole horizon from being eliminated by the
absorption of waves which may ``supersaturate" the extremality
condition, Eq. (\ref{Eq1}).

One may try to ``over spin" a black hole by sending into it waves
with large angular momentum. Recently, it has been claimed
\cite{MatSil} that this process may indeed push a near-extremal
black hole over the extremal limit. In particular, it has been shown
that a charged (Reissner-N\"ordstrom) black hole may acquire enough
angular momentum to over spin, $M^2-Q^2-a^2<0$. The authors of
\cite{MatSil} therefore concluded that such processes may serve as
counterexamples to the WCCC.

It is important to realize, however, that previous analyzes
\cite{MatSil} considered only the zeroth-order interaction between
the black hole and the incident wave. That is, the wave was assumed
to propagate on a {\it fixed} (unperturbed) Reissner-N\"ordstrom
background. As we shall show below, backreaction effects turn out to
be a crucial ingredient of the analysis. In particular, we shall
demonstrate that self-energy corrections must be taken into account
in order to preserve the black-hole integrity and to insure the
validity of Penrose's cosmic censorship conjecture.

We analyze now the process in which massless scalar particles with
angular momentum are beamed from far away towards a near extremal
black hole. As mentioned, it is essential to take higher-order
backreaction affects into account. As the field spirals into the
black hole it interacts with the black hole, so the horizon
generators start to rotate. This implies that, even if the initial
(``bare") black hole was a non-rotating one (as assumed in
\cite{MatSil}), the field would ``ignite" its rotation, such that
the propagation of the field itself is actually taking place on a
slowly rotating perturbed spacetime. However small, these
backreaction effects must be taken into account. We shall therefore
allow for a small rotation of the perturbed spacetime. (This should
be contrasted with the spherically symmetric unperturbed spacetime
assumed in \cite{MatSil}.)

The dynamics of a scalar field $\Psi$ in the rotating Kerr-Newman
spacetime is governed by the Teukolsky equation \cite{Teu,Dud}. One
may decompose the field as

\begin{equation}\label{Eq2}
\Psi_{lm}(t,r,\theta,\phi)=e^{im\phi}S_{lm}(\theta;a\omega)\psi_{lm}(r;a\omega)e^{-i\omega
t}\  ,
\end{equation}
where $(t,r,\theta,\phi)$ are the Boyer-Lindquist coordinates,
$\omega$ is the (conserved) frequency of the mode, $l$ is the
spheroidal harmonic index, and $m$ is the azimuthal harmonic index
with $-l\leq m\leq l$. (We shall henceforth omit the indices $l,m$
for brevity.) With the decomposition (\ref{Eq2}), $\psi$ and $S$
obey radial and angular equations, both of confluent Heun type
\cite{Heun,Flam}, coupled by a separation constant $A(a\omega)$. The
radial Teukolsky equation is given by

\begin{equation}\label{Eq3}
{{d}
\over{dr}}\Big(\Delta{{d\psi}\over{dr}}\Big)+\Big[{{K^2}\over{\Delta}}-(a\omega)^2+2ma\omega-A\Big]\psi=0\
,
\end{equation}
where $\Delta\equiv r^2-2Mr+Q^2+a^2$ and $K\equiv
(r^2+a^2)\omega-am$. The zeroes of $\Delta$, $r_{\pm}=M\pm
(M^2-Q^2-a^2)^{1/2}$, are the black hole (event and inner) horizons.
The functions $S(\theta;a\omega)$ are the spheroidal wave functions
\cite{Teu,Flam}. In the $a\omega\ll 1$ limit they become the
familiar spherical harmonics with the corresponding angular
eigenvalues $A_{lm}=l(l+1)+O(a\omega)$ \cite{Notealm}.

One should impose physical boundary conditions of purely ingoing
waves at the black-hole horizon and a mixture of both ingoing and
outgoing waves at infinity (these correspond to incident and
scattered waves, respectively). That is,

\begin{equation}\label{Eq4}
\psi \sim
\begin{cases}
e^{-i\omega y}+{\mathcal{R}}(\omega)e^{i \omega y} & \text{ as }
r\rightarrow\infty\ \ (y\rightarrow \infty)\ ; \\
{\mathcal{T}}(\omega)e^{-i (\omega-m\Omega)y} & \text{ as }
r\rightarrow r_+\ \ (y\rightarrow -\infty)\ ,
\end{cases}
\end{equation}
where the ``tortoise" radial coordinate $y$ is defined by
$dy=[(r^2+a^2)/\Delta]dr$. Here $\Omega$ is the angular velocity of
the black hole. The coefficients ${\cal T}(\omega)$ and ${\cal
R}(\omega)$ are the transmission and reflection amplitudes for a
wave incident from infinity.

The transmission and reflection amplitudes satisfy the usual
probability conservation equation $|{\cal T}(\omega)|^2+|{\cal
R}(\omega)|^2=1$. The calculation of these scattering amplitudes in
the low frequency limit, $M\omega\ll 1$, is a common practice in the
physics of black holes, see e.g. \cite{Chan,Page} and references
therein. Define

\begin{equation}\label{Eq5}
x\equiv {{r-r_+}\over {r_+-r_-}}\ \ ;\ \
\varpi\equiv{{\omega-m\Omega}\over{4\pi T_{BH}}}\ \ ;\ \ k\equiv
\omega(r_+-r_-)\  ,
\end{equation}
where $T_{BH}={{(r_+-r_-)}\over{4\pi(r^2_++a^2)}}$ is the
Bekenstein-Hawking temperature of the black hole. Then a solution of
Eq. (\ref{Eq3}) obeying the ingoing boundary conditions at the
horizon ($r\to r_+$, $kx\ll 1$) is given by \cite{Morse,Abram}

\begin{equation}\label{Eq6}
\psi=x^{-i\varpi}(x+1)^{i\varpi} {_2F_1}(-l,l+1;1-2i\varpi;-x)\  ,
\end{equation}
where $_2F_1(a,b;c;z)$ is the hypergeometric function. In the
asymptotic ($r\gg M$, $x\gg \varpi +1$) limit one finds the solution
\cite{Morse,Abram}

\begin{eqnarray}\label{Eq7}
\psi&=&C_1e^{-ikx}x^l{_1F_1}(l+1;2l+2;2ikx)\nonumber
\\&& +C_2e^{-ikx}x^{-(l+1)}{_1F_1}(-l;-2l;2ikx)\ ,
\end{eqnarray}
where $_1F_1(a;c;z)$ is the confluent hypergeometric function. The
coefficients $C_1$ and $C_2$ can be determined by matching the two
solutions in the overlap region $\varpi+1\ll x\ll 1/k$. This yields

\begin{eqnarray}\label{Eq8}
C_1&=&{{\Gamma(2l+1)\Gamma(1-2i\varpi)}\over{\Gamma(l+1)\Gamma(l+1-2i\varpi)}}\nonumber
\\
\text{and}\ \
C_2&=&{{\Gamma(-2l-1)\Gamma(1-2i\varpi)}\over{\Gamma(-l)\Gamma(-l-2i\varpi)}}\
.
\end{eqnarray}
Finally, the asymptotic form of the confluent hypergeometric
functions \cite{Morse,Abram} can be used to write the solution in
the form given by Eq. ({\ref{Eq4}). After some algebra one finds

\begin{eqnarray}\label{Eq9}
|{\cal
T}(\omega)|^2&=&\Big[{{l!^2}\over{(2l)!(2l+1)!!}}\Big]^2\prod_{n=1}^{l}\Big[1+\Big({{\omega-m\Omega}\over{2\pi
T_{BH}n}}\Big)^2\Big]\nonumber \\&& \Big({{\omega-m\Omega}\over{\pi
T_{BH}}}\Big)[(r_+-r_-)\omega]^{2l+1}\  ,
\end{eqnarray}
for the transmission probability.

To {\it zeroth} order in wave-hole interaction the wave propagates
on a fixed (unperturbed) black-hole background, where the zeroth
order angular velocity of the black hole is given by
$\Omega^{(0)}=a/(r^2_++a^2)$.

One should also consider {\it first}-order interactions between the
black hole and the angular momentum of the incident wave. As the
wave spirals into the black hole it interacts with the black hole,
so the horizon generators start to rotate, such that at the point of
absorption the black-hole angular velocity has changed from
$\Omega^{(0)}$ to $\Omega^{(0)}+\Omega^{(1)}$. On dimensional
analysis one expects $\Omega^{(1)}$ to be of the order of
$O(m/M^3)$. In fact, Will \cite{Will} has performed a perturbation
analysis for the problem of a ring of particles rotating around a
slowly spinning (neutral) black hole, and found
$\Omega^{(1)}=m/4M^3$. For a charged black hole one finds a similar
result,

\begin{equation}\label{Eq10}
\Omega^{(1)}={{m}\over{Mr^2_+}}\  .
\end{equation}
(We note that this last result reduces to $\Omega^{(1)}=m/4M^3$ in
the neutral case, as found in \cite{Will}). As would be expected
from a perturbative approach, $\Omega^{(1)}$ is proportional to $m$
\cite{simple}.

Taking cognizance of the transmission probability, Eq. (\ref{Eq9}),
one realizes that those modes for which the frequency $\omega$ and
the azimuthal quantum number $m$ are related by $\omega<m\Omega$
have negative transmission probabilities. These modes are actually
amplified rather than absorbed. This is the well-known black-hole
superradiance phenomena \cite{PreTeu,Beko}. Thus, only modes for
which

\begin{equation}\label{Eq11}
\omega>m\Omega\  ,
\end{equation}
can be absorbed by the black hole.

If the original (unperturbed) black hole was a near extremal
Reissner-N\"ordstrom one (as assumed in \cite{MatSil}) then
$\Omega^{(0)}=0$. Substituting Eq. (\ref{Eq10}) into Eq.
(\ref{Eq11}), one finds the absorption condition

\begin{equation}\label{Eq12}
\omega>m^2/M^3\  .
\end{equation}
Only modes which respect the inequality (\ref{Eq12}) can be absorbed
by the black hole.

Our aim is to challenge the validity of the WCCC in the most
``dangerous" situation, i.e., when the energy delivered to the black
hole is as small as possible. We shall therefore consider a single
mode of azimuthal angular momentum $m$ and energy $\omega=m^2/M^3$.
The absorption of the mode by the black hole produces the following
changes in the black-hole parameters: $M\to M+\omega$ and $a=0\to
m/M$. Hence, the condition (\ref{Eq1}) for the black hole to
preserve its integrity after the absorption of the mode is now given
by

\begin{equation}\label{Eq13}
(M+m^2/M^3)^2-Q^2-(m/M)^2\geq 0\  .
\end{equation}
Since the parameters of the original black hole conformed to the
relation $M^2-Q^2\geq 0$, we find that the black-hole condition
(\ref{Eq13}) is indeed satisfied \cite{Noteback}. Thus one concludes
that the incident mode cannot remove the black-hole horizon. Cosmic
censorship is therefore respected!

It is worth reexamining a similar gedanken experiment which has been
designed to challenge cosmic censorship. One may try to over charge
a near extremal Reissner-N\"ordstrom black hole by dropping into it
a charged particle. Neglecting backreaction effects (that is,
restricting ourselves to the test particle approximation), one
concludes \cite{Hub} that the charged particle may over charge the
black hole. However, when backreaction effects are partially taken
into account \cite{Hub}, one finds that the particle may actually
bounce back before reaching the black-hole horizon. This implies
that the corrected process actually fails to destroy the black hole.

It should be mentioned, however, that the perturbation analysis
presented in \cite{Hub} was a numerical one. As such, the outcome of
this numerical analysis can not be regarded as a generic one. It is
therefore desirable to calculate the backreaction effects
analytically, as we shall do below.

The total energy of a charged particle of mass $\mu$ and charge $q$
in the black-hole spacetime is made up of three contributions: $1)$
${\cal E}_{0}=\mu (g_{00})^{1/2}$, the energy associated with the
particle's mass (red-shifted by the gravitational field); $2)$
${\cal E}_{elec}=qQ/r$, the electrostatic interaction of the charged
particle with the black-hole electric field; and $3)$ ${\cal
E}_{self}$, the gravitationally induced self-energy of the charged
particle.

The third contribution ${\cal E}_{self}$ reflects the effect of the
spacetime curvature on the particle's electrostatic
self-interaction. The physical origin of this force is the
distortion of the charge's long-range Coulomb field by the spacetime
curvature. This can also be interpreted as being due to the image
charge induced inside the (polarized) black hole
\cite{Linet,BekMay}. The self-interaction of a charged particle in
the black-hole spacetime results with a repulsive (i.e., directed
away from the black hole) self-force. A variety of techniques have
been used to demonstrate this effect in the black-hole spacetime. In
particular, one finds \cite{Zel,Loh} ${\cal E}_{self}=Mq^2/2r^2$ in
the Reissner-N\"ordstrom spacetime.

The absorption of the injected particle by the black hole produces
the following changes in the black-hole parameters (assuming that
the energy delivered to the black hole is as small as possible):
$M\to M+qQ/r_++Mq^2/2r^2_+$ and $Q\to Q+q$. Hence, the condition
(\ref{Eq1}) for the black hole to preserve its integrity after the
absorption of the charged particle is given simply by

\begin{equation}\label{Eq14}
(q-\epsilon)^2\geq 0\  ,
\end{equation}
where $r_{\pm} \equiv M {\pm} \epsilon$. This condition is satisfied
trivially \cite{Noteback2}, and we therefore recover our previous
conclusion-- cosmic censorship is respected!

In summary, extreme situations which have been considered as
counterexamples to the weak cosmic censorship conjecture were
reexamined. In particular, we have reanalyzed gedanken experiments
in which scalar waves carrying large angular momentum are absorbed
by a black hole. At first sight, it seems that the black hole may
acquire enough angular momentum to over spin, $M^2-Q^2-a^2<0$.
Previous analyzes \cite{MatSil} indeed claimed that this process may
provide a counterexample to the WCCC. However, we have demonstrated
that a more complete analysis of the gedanken experiments (in which
backreaction effects are properly taken into account) reveals that
they do {\it not} violate the weak cosmic censorship conjecture.
This teaches us that backreaction effects must be taken into account
in order to secure the black-hole integrity and to affirm the
validity of the cosmic censorship conjecture. It is worth
emphasizing that saving cosmic censorship in such extreme situations
is essential for preserving the predictability of Einstein's theory
of gravity.

\bigskip
\noindent
{\bf ACKNOWLEDGMENTS}
\bigskip

This research is supported by the Meltzer Science Foundation. I
thank Jacob D. Bekenstein, Tsvi Piran and Uri Keshet for stimulated
discussions. I also thank Yael Oren for helpful discussions.



\begin{thebibliography}{99}
\bibitem{Notealm} In the small $a\omega$ limit we shall take
$A=l(l+1)$ in Eq. (\ref{Eq3}), where $l$ is nearly an integer with a
small correction of order $O(a\omega)$.
\bibitem{simple} These backreaction effects could
also be inferred from the following heuristic argument: assume that
the mode of angular index $m$ has induced a change in the spacetime
parameters, such that the specific angular momentum of the spacetime
has increased from $0$ to $m/M$. Substituting this perturbed value
into the usual expression $\Omega=a/(r^2_++a^2)$, one indeed finds
$\Omega={{m}\over{Mr^2_+}}+O(m^2/M^4)$.
\bibitem{Noteback} It should be emphasized that without backreaction
effects, the black-hole condition, Eq. (\ref{Eq13}), would become
$M^2-Q^2-(m/M)^2\geq 0$. One therefore realizes that without the
backreaction effects it would have been possible to violate the
black-hole condition by sending a mode with large enough angular
momentum.
\bibitem{Noteback2} It is worth emphasizing that without backreaction
effects (namely, the black-hole polarization), the black-hole
condition, Eq. (\ref{Eq14}), would become $\epsilon^2-2q\epsilon
\geq 0$. One therefore realizes that without the backreaction
effects it would have been possible to over charge the black hole by
sending into it a charged (test) particle with $q>\epsilon/2$.
\end{thebibliography}
\end{document}